# Dynamics of interaction of topological localized structures in reversed time

Farhod Shokir

*S.U.Umarov Physical -Technical Institute*
*of the National Academy of Sciences of Tajikistan*

**Abstract:** The results of numerical simulation of the interaction of topological solitons (2+1)-dimensional O(3) non-linear sigma model in reversed time are presented. At the first stage, models of interactions of topological vortices are developed, where, depending on the dynamic parameters, processes of their decay into localized perturbations and phased annihilation are observed. Also, models for the phased annihilation of topological vortices during their interaction with 180-degree domain walls are considered. On the basis of the models obtained, initial conditions for numerical simulation of interaction processes in reversed time are developed. The models describing the complete restoration of the initial topological field of interacting solitons at the combination of localized perturbations and radiation waves are obtained. Also, models are obtained that describe the formation of topological vortices in the plane of the domain wall and their subsequent emission. Thus, the T-invariance property of the field-theoretic model under study is confirmed. Numerical calculations were carried out in a stratified space on the basis of methods of the theory of finite difference schemes, using the properties of a stereo-graphic projection. The experiments were carried out for different values of the Kronecker-Hopf index of topological vortices. A complex program module is proposed that implements a special algorithm for the numerical calculation of the evolution of the interaction of space-time topological structures in reversed time.

## 1. Introduction

The most general characteristic of physical laws is their symmetry corresponding to conservation laws, while the concept of symmetry has a broader and deeper meaning than, for example, in ordinary geometry [1]. Consider the T-invariance property of a (2+1)-dimensional supersymmetric O(3) non-linear sigma model (NSM). First proposed at the turn of the 1950s and 1960s to effectively describe the nature of massless excitations, the NSM theory is used as a reliable theoretical basis for research under simplified conditions. As was first noted by A.M. Polyakov [2], there is a deeply rooted analogy between four-dimensional Yang-Mills theories and two-dimensional NSM. Thus, NSM are theoretical laboratories for approbation of methods and approaches developed for problems of real physics [3].

The study of the properties of the T-symmetry of physical phenomena are of great importance in applied problems (see, for example, [4–12]), including medical imaging, echo-pulse control, the development of nanophotonic devices, switchable narrow-band optical isolators, and also in underwater acoustics and in the study of extreme waves. Paper [4] reports the first experiments showing the reversibility of multiply scattered acoustic waves. It was found that when modeling processes in reversed time ($t' \rightarrow -t$), the waves converge to their source and restore their original shape. In [5], the time reversal (reversed time – RT) method was applied to recover strongly localized perturbations (extreme waves) of dispersive media in the framework of the nonlinear Schrödinger equation. It is shown that highly localized nonlinear waves can be experimentally reconstructed by the RT method in other nonlinear dispersive media controlled by the nonlinear Schrödinger equation, such as optics, plasma and Bose-Einstein condensate. A method for reproducing dispersion-insensitive Hong–U–Mandela interferograms based on the properties of the T-symmetry of quantum mechanics was proposed in [6]. In [7], a quasi-two-dimensional scheme of a two-layer system of fermions immersed in a Bose-Einstein condensate was proposed for the implementation of T-invariant topological superfluids. The results of this work show that the experimental realization of a topological superfluid liquid with T-symmetry can be carried out in a cold atomic system. Practically important results for the development of nonlinear insulators, metasurfaces, and other nanophotonic devices are presented in [8]. The works [9, 10] are devoted to the study of certain quantities (indicators of anomalies) that determine the anomaly of T-invariant topologically ordered states. The anomaly in this case has the meaning of the impossibility of realizing the above states in strictly two-dimensional systems. Interesting research in the development of so-called photonic devices for use in quantum materials and processors was carried out in [11]. Also noteworthy is the work [12], where the authors describe a method for developing highly sensitive sensors for recording high-energy physics processes outside the framework of the Standard Model.

The above review of works reflects only a part of theoretical studies and practical experiments carried out on the basis of T-invariant properties of physical processes. Applications of RT methods in modeling seismological and geophysical processes, in nondestructive testing of solid materials [13], in the study of high-temperature superconductors, etc. are not mentioned. The study of the properties of the T-invariance O(3) of the NSM is carried out by considering the processes of interaction, decay, and annihilation of its topological solutions. Since the TR operation ($t' \to -t$) is a transposition of the initial and final states of the system, its action does not lead to new characteristics of the studied quasiparticles – topological localized structures (topological solitons – TS). Recall that the density of the Lagrange function and the Hamiltonian of the studied O(3) NSM in the standard (isospin) parametrization can be written in the following form (see, for example, [14–18]):

$$\mathcal{L} = g[\partial_\mu s_a \partial^\mu s_a + (s_3^2 - 1)] \tag{1}$$

$$\mathcal{H} = g[(\partial_0 s_a)^2 + (\partial_1 s_a)^2 + (\partial_2 s_a)^2 + 1 - s_3^2] \tag{2}$$

where $g = 1/2; \mu = 0, 1, 2; a = 1, 2, 3; s_a s_a = 1$ – coordinates of a single isovector **S** (triplet of real scalar fields)

$$\mathbf{S} = \begin{pmatrix} s_1^{(1/2)} \\ s_2^{(1/2)} \\ s_3^{(1/2)} \end{pmatrix} = \begin{pmatrix} \sin\theta_{1/2} \cos\varphi_{1/2} \\ \sin\theta_{1/2} \sin\varphi_{1/2} \\ \cos\theta_{1/2} \end{pmatrix} = \begin{pmatrix} \frac{2 x_{1/2}}{1+\zeta_{1/2}\zeta_{1/2}^*} \\ \frac{2 y_{1/2}}{1+\zeta_{1/2}\zeta_{1/2}^*} \\ \frac{1-\zeta_{1/2}\zeta_{1/2}^*}{1+\zeta_{1/2}\zeta_{1/2}^*} \end{pmatrix} \tag{3}$$

In this case, a restriction is imposed and the length of the isovector (3) is a constant value; $\theta(t)$ and $\varphi(t)$ – Euler angles (Fig. 1a). O(N) NSM in two-dimensional space-time is the theory of n-fields of $\sigma^i$ ($i = 1, ..., n$) defined on the unit sphere [3]. Thus, the n-field of the studied model (1) is described by the motion of a point on the Bloch sphere ($S^2$)

$$S^2 = SU(2)/U(1) = SO(3)/SO(2), \qquad (4)$$

equivalent to the movement of the end of the isovector **S** (Fig. 1a). Moreover, there is a one-to-one correspondence between the sphere and factor groups SU(2)/U(1) (of the special unitary group SU(2) with respect to the unitary group U(1)) and (of the special orthogonal group SO(3) with respect to the group SO(2)).

The first term in (1) is the known NSM Lagrange function for the isotropic case [16, 18]. The study was carried out within the framework of the anisotropic O(3) NSM, where the anisotropy is chosen in the direction of the $s_3$-component. Thus, the states of the studied model (1) with zero energy (vacuum states) are described in the layered space (Fig. 1b) by the isovector $\mathbf{S}(0,0,\pm 1)$. The symmetry of the model is O(3), which is the symmetry of the dynamics of the isovector in the sphere (4).

The Lagrange function O(3) NSM does not explicitly depend on time and contains only even degrees of derivatives, so the application of the operation T should not lead to a change in the equation of motion. Consequently, each physical state obtained in the framework of O(3) NSM at the moment of time $t$ at RT ($t' \to -t$) passes into a T-invariant state with reversed directions of velocities. It is known that the passage of time is aimed at the implementation of processes with the highest probability, but physical laws do not prohibit the implementation of also unlikely processes.

Thus, the property of T-invariance O(3) NSM allows the existence of the last state, as well as time-reversed motion. Using the example of constructing numerical models for the interaction of quasiparticles – topological vortices and domain walls (DW) in RT ($t' \to -t$), the T-symmetry property of the studied sigma model is directly confirmed. A set of computer programs based on a special algorithm for modeling the processes of interaction of particle-like solutions of nonlinear equations of field theory models in RT has been developed.

In [19], in the framework of O(3) NSM, models of two-soliton interactions of topological solitons-vortices of the following form were obtained [20]:

$$\theta(r,R) = 2\text{arctg}\left(\frac{r}{R}\right)^{Q_t}, \quad \varphi = Q_t\chi - \omega\tau, \tag{5}$$

$$r^2 = x^2 + y^2, \quad \cos\chi = \frac{x}{r}, \quad \sin\chi = \frac{y}{r},$$

where $Q_t$ – topological charge [1–3, 20–22] (mapping degree, Kronecker-Hopf index); $\chi$ – is the angular parameter; $r$ – is the soliton localization radius; is the frequency of rotation of the isovector in the space of the sphere $S^2$ (4); $R$ – variational parameter; $\tau$ – time step.

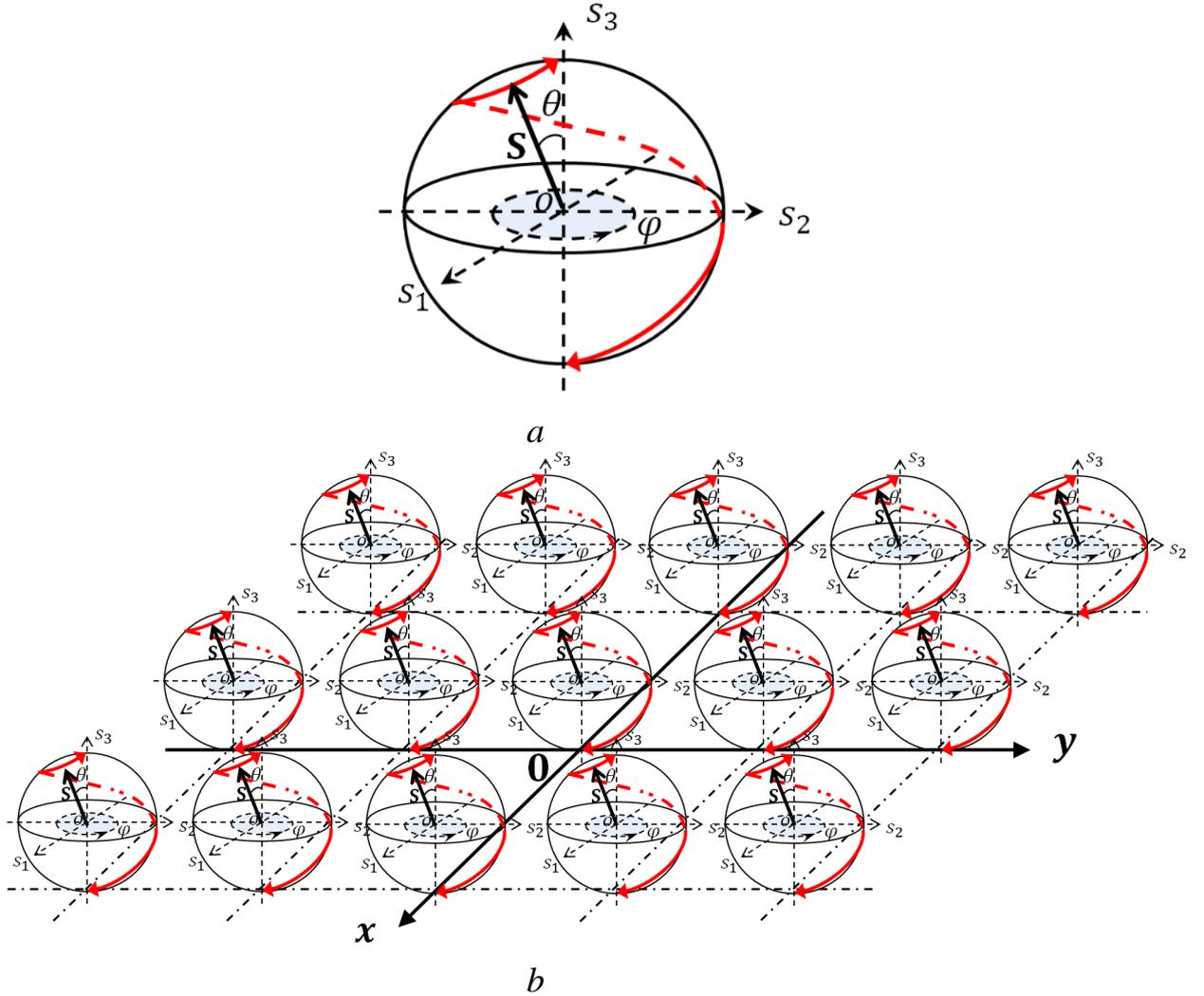

**Figure 1.** Bloch sphere – $S^2$ (a); stratified space – $2D$ (b)

On Fig. 2 shows illustrations of the energy density (a), isovector structure (b) and its projection onto the plane (c) for TS (5) O(3) NSM with topological charges $Q_t = 1, ..., 4$. All illustrations are given in dimensionless quantities, since the O(3)

NSM equations used in these experiments are of a universal nature, regardless of applications to various fields of physics. It should also be noted that in mathematical modeling it is dimensionless quantities that can reveal the complex nature and general properties of the processes under study. In the experiments performed, the only physically measurable quantity is the velocity of solitons ($\mathbf{v}(t_0)$) specified by the Lorentz transformations. In this case, for example, for problems of quantum field theory, Planck units of measurement can be used, but, for problems in other areas of physics, it is advisable to use the corresponding derived systems of units. Thus, the developed research methods and the resulting models can be used by specialists in the field, for example, elementary particles or other areas of physics, and specific values are calculated based on the calculated data.

At $Q_t = 1$ (skyrmion [16]), the energy density (DH) of the vortex (5) is concentrated in the central part of the simulation area $L[1001 \times 1001]$. At $Q_t \geq 2$ DH (2) of topological vortices (5) are concentrated in an annular shape (Fig. 2a), which is a domain boundary between opposite directions of the field ($\pm s_3$). An increase in the values of $Q_t$ leads to a narrowing of the width of this domain wall and an increase in its energy density. From Fig. 2b it can be seen that for each case of $Q_t = 1, \ldots, 4$ in the isovector structure of TS (5) there are vortex structures, the so-called local vorticity of the field [22], the number of which is equal to the $Q_t$ value. In the central part of the TS (as well as in the centers of the vortex structures), the vectors of the isotopic spin $\mathbf{S}$ are co-directed with the positive direction of the axis $Z$: $s_3 \to 1$. At the edges of the layered space modeling area (Fig. 1b), the isovectors are directed in the opposite direction: $s_3 \to -1$. The transition between these states occurs by isovector $\mathbf{S}$ precession. Thus, TS (5) are topological perturbations of the field connecting vacuum states $\theta(t) = 0, \pi$ (Fig. 1a).

For numerical experiments, we used TS (5) with Hopf index values of $Q_t = 1, \ldots, 4$. In this case, the following expressions are valid for the coordinates of a unit isovector $\mathbf{S}$:

$$S_{Q_t=1} = \lambda_1^+ \begin{pmatrix} \xi_1^1 \cos\tau + \xi_1^2 \sin\tau \\ \xi_1^2 \cos\tau - \xi_1^1 \sin\tau \\ \lambda_q^- \end{pmatrix}, \quad S_{Q_t=2} = \lambda_2^+ \begin{pmatrix} \xi_2^1 \cos\tau - \xi_2^2 \sin\tau \\ \xi_2^2 \cos\tau - \xi_2^1 \sin\tau \\ \lambda_q^- \end{pmatrix},$$

$$S_{Q_t=3} = \lambda_3^+ \begin{pmatrix} \xi_3^1 \cos\tau - \xi_3^2 \sin\tau \\ -\xi_3^2 \cos\tau - \xi_3^1 \sin\tau \\ \lambda_q^- \end{pmatrix}, \quad S_{Q_t=4} = \lambda_4^+ \begin{pmatrix} \xi_4^1 \cos\tau + \xi_4^2 \sin\tau \\ \xi_4^2 \cos\tau - \xi_4^1 \sin\tau \\ \lambda_q^- \end{pmatrix},$$

where

$$\lambda_q^+ = 2(1 + r^{2q})^{-1}, \; \lambda_q^- = 2^{-1}(1 - r^{2q}), \; q = 1, \ldots, 4,$$

$$\xi_1^1 = x, \; \xi_1^2 = y, \; \xi_2^1 = x^2 - y^2, \; \xi_2^2 = 2xy,$$

$$\xi_3^1 = x^3 - 3xy^2, \; \xi_3^2 = y^3 - 3x^2y,$$

$$\xi_4^1 = x^4 - 6x^2y^2 + y^4, \; \xi_4^2 = 4x^3y - 4xy^3.$$

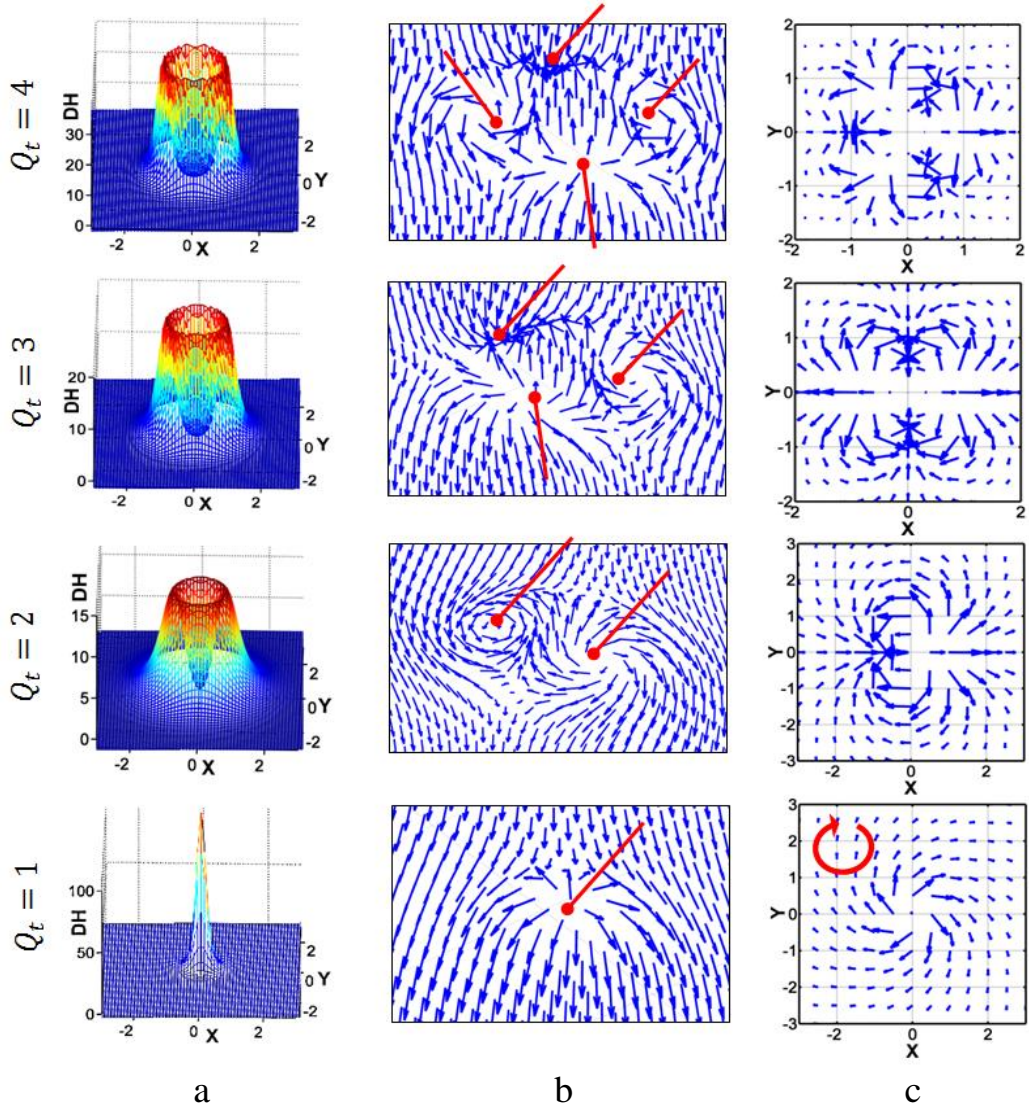

a          b          c

**Figure 2.** DH of TS (a); TS isovector structure – 3D (b); TS isovector structure – 2D c).

In [19], models of the decay of interacting TS of a (2+1)-dimensional O(3) NSM into localized perturbations were obtained, the topological charge sum conservation property was revealed, and a method was proposed for determining the topological charge of localized perturbations. In [23], the isospin dynamics of solutions (5) was studied and the conditions for the manifestation of long-range forces during their interaction, as well as complete annihilation by staged energy emission, were determined. It was shown in [23] that when TS (5) collides with a TS of the following form:

$$\theta(r, R) = 2\operatorname{arctg}\left(\frac{r}{R}\right)^{Q_t}, \quad \varphi = Q_t \chi - \omega \tau, \qquad (6)$$

$$r^2 = x^2 + y^2, \qquad \cos \chi = -\frac{x}{r}, \qquad \sin \chi = \frac{y}{r},$$

(which differs in the sign of the cosine of the angular parameter χ), the process of their gradual annihilation is observed. At the same time, at each stage, energy equivalent to a unit of topological charge ($Q_t = 1$) is emitted in the form of a pair of linear perturbation waves (with $Q_t = \frac{1}{2}$) propagating similarly to the previous case with a speed $c = 1$. As shown in [23], the condition for the stage-by-stage annihilation of the ES of the form (5) is the codirection ($S(\uparrow\uparrow)$) and synchronous correlated motion ($S_{cor}$) of the isovectors $S$ (3) of the interacting TS in the collision area: $S(\uparrow\uparrow) \cap S_{cor}$. In this case, the process of phased annihilation occurs at any speed of the vehicle. Note also that the long-range interaction condition for two-soliton interactions of TS of the form (5) is $S(\uparrow\downarrow) \cap S_{cor}$ [23].

In [24], in the framework of O(3) NSM, models of the interaction of topological vortices (5) with known topological solutions of the sine-Gordon equation in the form of a DW were obtained

$$\operatorname{tg}\frac{\theta}{2} = e^{B_1 \frac{w}{k_1}(x-x_0) + B_2 \frac{w}{k_2}(y-y_0)}, \quad \varphi(t) = \epsilon. \qquad (7)$$

Note that solution (7) for $\epsilon = 0, \pi$ and $\epsilon = \frac{\pi}{2}, \frac{3\pi}{2}$ describes the dynamics of the so-called Neel (N) and Bloch (B) DW, respectively

$$S_{N(0,\pi)} = \Lambda^{-1}\begin{pmatrix}\pm 2e^x \\ 0 \\ 1-e^{2x}\end{pmatrix}, \qquad S_{B\left(\frac{\pi}{2},\frac{3\pi}{2}\right)} = \Lambda^{-1}\begin{pmatrix}0 \\ \pm 2e^x \\ 1-e^{2x}\end{pmatrix},$$

where $\Lambda = 1 + e^{2x}$. In this case, the isospin parameters $s_i$ ($s_i s_i = 1$, $i = 1, 2, 3$) correspond to the coordinates of the unit isovector (3). In [24], a numerical study of models of two-soliton interactions of topological solutions of the form (5) and (6) (for $\epsilon = 0$) was carried out. Results of each series ($Q_t = 1, \ldots, 6$; $\epsilon = 0, \frac{\pi}{2}, \pi, \frac{3\pi}{2}$) of experiments in the above work can be characterized in a unified way - the gradual decay of topological vortices (5) into localized perturbations with half ($Q_t = \frac{1}{2}$) values of the topological charge. At each stage of the decay processes, radiation of the energy of a topological vortex is observed, equivalent to a unit topological charge ($Q_t = 1$), in the form of two localized perturbations moving along the DS plane at a speed $c = 1$ (the speed of light in vacuum).

As noted above, the RT method can be applied to every phenomenon described by equations that contain only derivatives of an even order with respect to time. For each solution $z(r, t)$ there is a second solution of the form $z(r, -t)$, since their second derivatives coincide. Another limitation for the present approach requires that the medium be non-dissipative. To implement this condition, it is necessary to restore the radiation of energy. In the experiments under consideration, in models with RT, radiation waves are reconstructed within the simulation area $L(x, y)$. But the recovery of waves absorbed by the boundary conditions ($En_{loss}$) was not provided, since they are equivalent to a negligibly small part of the total energy of the models under study: $En_{loss} \in (10^{-3}, 10^{-2})$.

Thus, we have considered the problem of constructing models that describe the time-reverse (T-invariant) evolution of the processes of (two-soliton) frontal collision, decay, and annihilation of the TS of the form (5), (6), and (7). The approach of approximating difference schemes [25] with the use of the algorithm and numerical scheme proposed in [17] for modeling the stationary TS (5) within the O(3) NSM was used. The experimental results confirm the T-invariance property of O(3) NSM. The developed algorithms, numerical schemes and computer codes are combined into a single package of applied programs that allow researching T-invariant properties of (2 + 1)-dimensional field theory models of the O(3) NSM class.

Next, we present the models obtained during RT, where the structure of their initial states is restored with a sufficiently high degree of accuracy.

**2. Interaction and decay of TS in reversed time**

As was shown in our previous studies (see, for example, [18, 19, 26]), when simulating two-soliton head-on collisions of TS (5) moving at relatively low velocities, the processes of their collision and mutual reflection are observed. But, with an increase in the speed of movement of the interacting vortices, they disintegrate into localized perturbations (LP) that scatter away from the point of collision. At the same time, LP also have topological charges: $Q_t(\text{LP}) < Q_t(\text{TS})$ A common property of the processes of TS decay into LP is the preservation of the sum of $Q_t$ values [18, 19]: $\sum Q_t(\text{LP}) \equiv Q_t(\text{TS}_1) + Q_t(\text{TS}_2)$.

On fig. 3a shows one of the examples of the implementation of the numerical model of the frontal collision of vehicles (5), which have equal values of $Q_t = 3$ and move in opposite directions with a speed $v(t_0) \approx \pm 0.287$. For smaller values of $v(t_0)$ the energy of the interacting TS is not sufficient for the implementation of their decay processes. On fig. 3a describes the process of head-on collision and decay of these vortices onto LP, which also have a topological charge: $Q_t = 1$ (two LP) and $Q_t = 1$ (two LP). All illustrations of this type show the values of the energy density of vortices – DH and its contour projection on the two-dimensional modeling area $L(2002 \times 1001)$. In this case, TS (5) are absolutely identical quasiparticles and have the property of chirality [18], so the process of interaction and expansion of the LP occurs along symmetrical trajectories.

At the second stage of the experiments, the final states (in the case of Fig. 3a: $t'_{0\pm\tau} = 38.4 \pm \tau$) of the obtained TS decay model were used as initial data for research in the RT ($t' \to -t$). On Fig. 3b shows illustrations describing the results of numerical simulation of the decay processes of TS (5) described in Fig. 3 in RT. At $t' \geq 14.4$, the LP move towards the center of the collision area, where they combine and form a single field disturbance. Further, there is a division of the formed single field perturbation into two ES ($t' \approx 23.4$) moving in $\pm x$-directions ($t' \to 38.4$).

These illustrations show that the model of the processes of collision and decay of TS (5) (Fig. 3b) obtained in the RT is symmetrical with respect to the original model (Fig. 3a) with a sufficiently high accuracy. Meanwhile, in the case of Fig. 3b, there is an insignificant energy loss of the system relative to the original model (Fig. 3a): $\text{En}_{loss} \approx 1.12\%$.

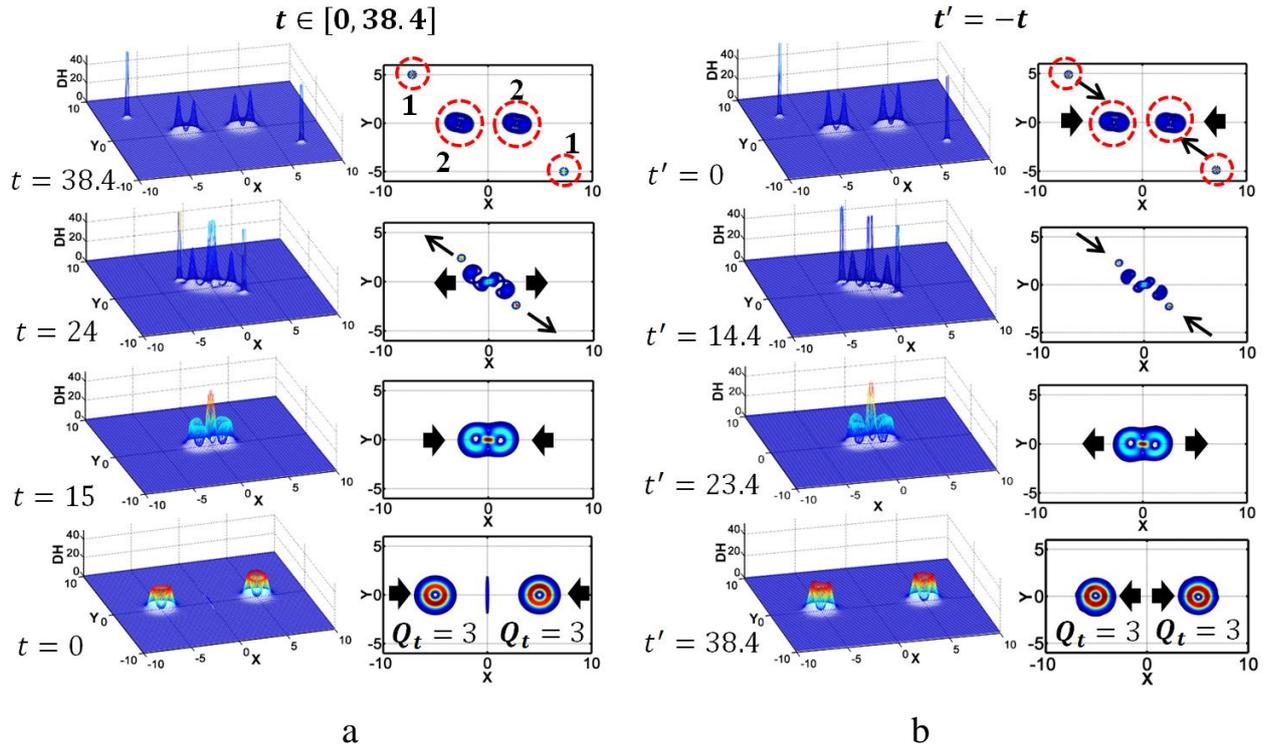

**Figure 3.** The energy density of the DH (and its projection onto the plane $z(x,y)$) of the processes of interaction and decay of two TS (5) at $Q_{t_{12}} = 3$: in ordinary – $t$ (a) and reversed time – ($t' \to -t$) (b).

On Fig. 4, a similar to the previous example, a model of a frontal collision of the vehicle (5) with $Q_t = 4$ at $v(t_0) \approx \pm 0.196$ is shown. Both vehicles are observed to break up during a collision into 3 pairs of LP: with $Q_t = 1$ (4 LP) and with $Q_t = 2$ (2 LP). Similar to the previous case, this process was simulated in the RT: $t'_{0\pm\tau} = 54 \pm \tau$ (Fig. 4b). In this case, the association of 6 separate LP into a single field disturbance is also observed, followed by the formation of two TS, which, separating, begin to move in $\pm x$ -directions.

These illustrations show that the model of collision and decay of TS (5) obtained in the RT (Fig. 4b) is also symmetrical with respect to the original model with a sufficiently high accuracy (Fig. 4a). In this case, the loss of the total energy of the system to radiation is equal to $\text{En}_{loss} \approx 0.91\%$.

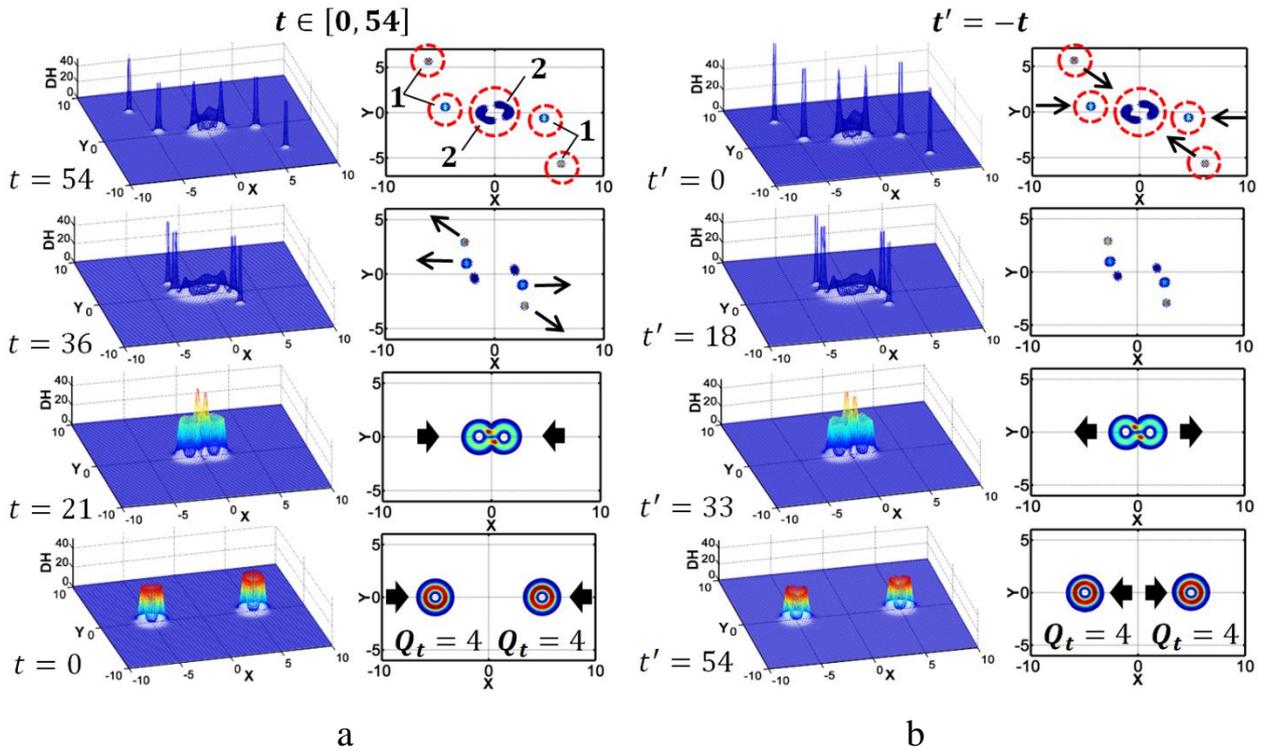

a          b

**Figure 4.** The energy density of the DH (and its projection onto the plane $z(x, y)$) of the processes of interaction and decay of two TS (5) at $Q_{t_{12}} = 4$: in ordinary – $t$ (a) and reversed time – $(t' \to -t)$ (b).

### 3. Phased annihilation of the TS in reversed time

On Fig. 5 shows the results of studies of the properties of T-invariance of the processes of interaction of TS (5) and (6) with $Q_t = 3$ and moving in opposite directions with a speed $\boldsymbol{v}(t_0) \approx \pm 0.196$. On Fig. 5a shows a model of a head-on collision and annihilation of these vortices obtained in the $L[2002 \times 3001]$ grid.

All illustrations of this type show the DH values of the interacting TS and its contour projection onto the two-dimensional area $L$. In this case, there is a gradual annihilation of the TS by periodic energy emission in the form of 3 pairs of LP waves.

At the second stage of experiments, the final states (in the case of Fig. 5a: $t'_{0\pm\tau} = 31.2 \pm \tau$) of the obtained model were used as initial data for studying the processes of TS annihilation in RT ($t' \to -t$).

On Fig. 5b shows illustrations of the obtained results. At $t' \approx 5.4$, the first pair of waves of the frontal part, interacting, form a well-localized bound state in the form of two TS, each of which has $Q_t = 1$. A similar process occurs at $t' \approx 8.4$ and $t' \approx 12.6$, thus, the values of the topological charges of both TS are restored to the original $Q_t = 3$. Further, the formed TS are separated and begin to move in $\pm x$ - directions. These illustrations show that the model of the processes of collision and annihilation of TS (5) and (6) (Fig. 5b) obtained in the RT is T-symmetric with respect to the original model (Fig. 5a). Meanwhile, in the case of Fig. 5b, the evolution of the system occurs practically without energy loss: $En_{loss} \to 0$.

On Fig. 6, a similar to the previous example, a model of a head-on collision and phased annihilation of TS (5) and (6) in the case of $Q_t = 4$ ($\boldsymbol{v}(t_0) \approx \pm 0.196$) is shown. The process of stage-by-stage annihilation of both TS is accompanied by the emission of disturbance waves in 4 periods. Similar to the previous case, this process was simulated in the RT: $t'_{0\pm\tau} = 31.5 \pm \tau$ (Fig. 5b).

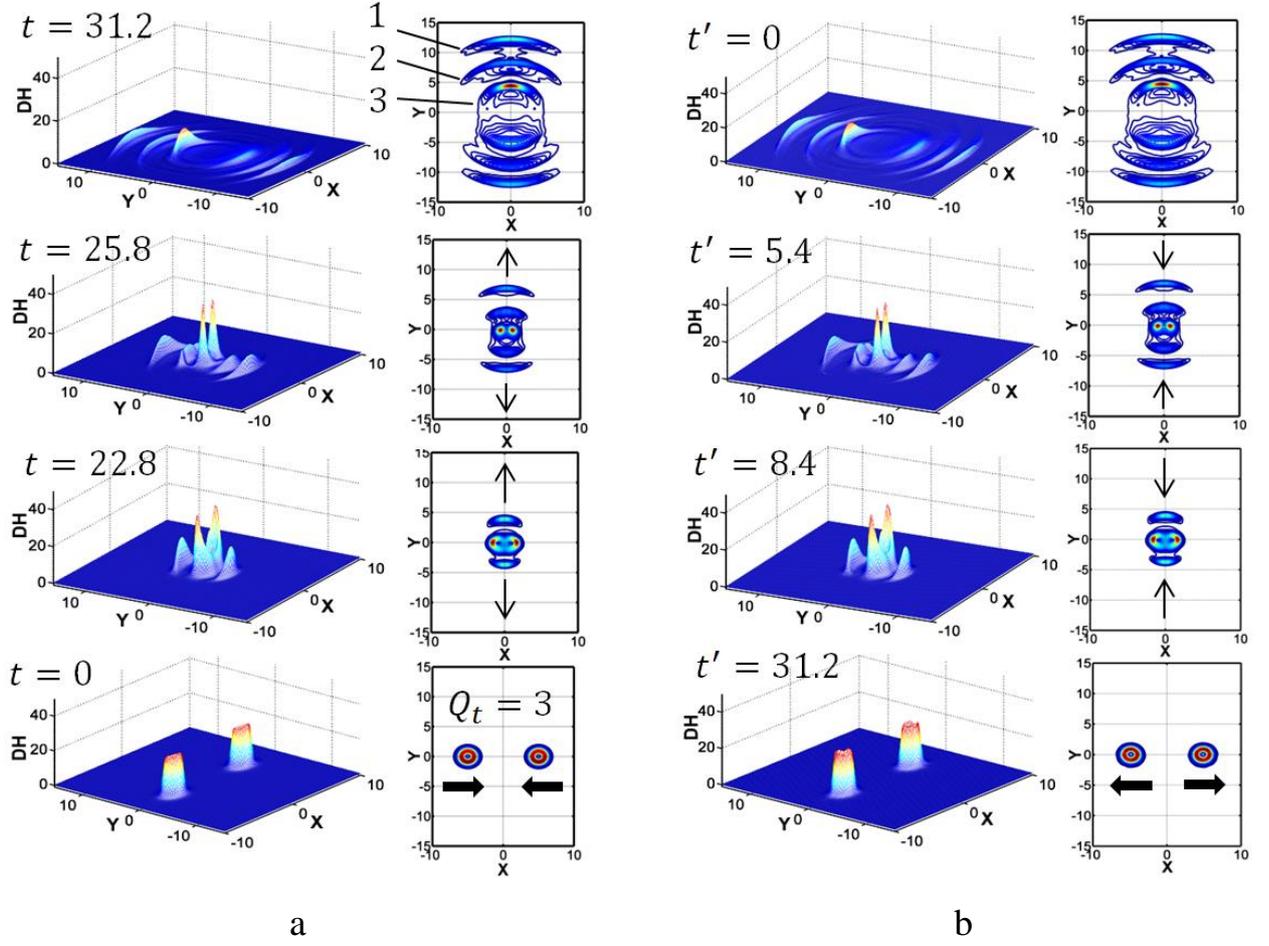

**Figure 5.** The energy density of the DH (and its projection onto the plane $z(x,y)$) of the processes of collision and annihilation of two TS (5) at $Q_{t_{12}} = 3$: in ordinary – $t$ (a) and reversed time – $(t' \to -t)$ (b).

In this case, there is also a concentration of radiation waves in the collision area and the gradual formation of two vehicles with $Q_t = 4$, which, separating, begin to move in $\pm x$ -directions. Thus, the model obtained in the RT is also T-symmetric with respect to the original model with a sufficiently high accuracy (Fig. 6a). In this case, the energy loss of the system of interacting vehicles in both cases is insignificant $\text{En}_{loss} \to 0$.

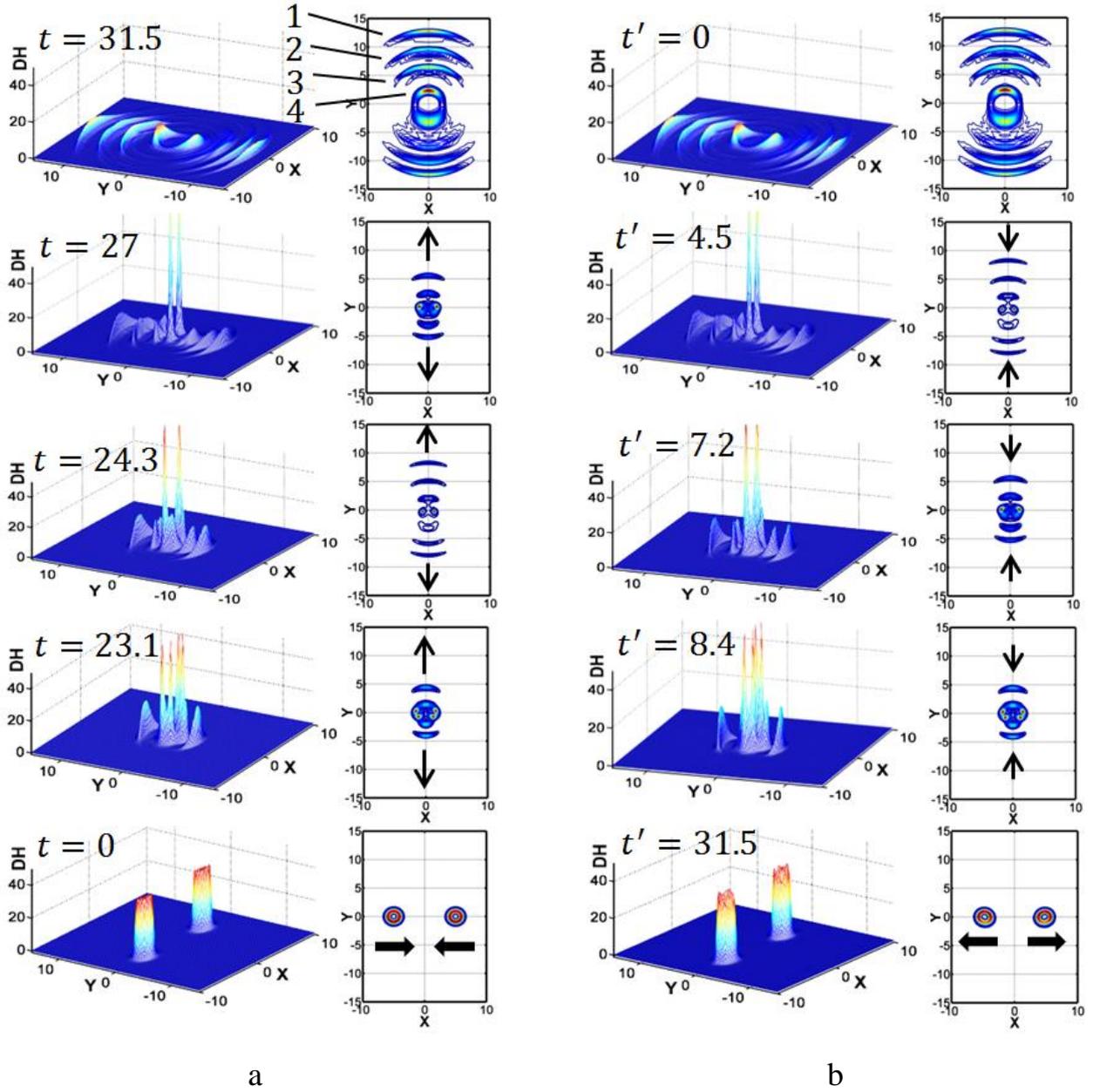

a  b

**Figure 6.** The energy density of the DH (and its projection onto the plane $z(x,y)$) of the processes of collision and annihilation of two TS (5) at $Q_{t_{12}} = 4$: in ordinary – $t$ (a) and reversed time – $(t' \to -t)$ (b).

## 4. Interaction of a TS with a DW in reversed time

As noted above, in our previous studies (see, for example, [24]), models of the decay of topological vortices (5) in the DW plane of the form (7) were obtained. On Fig. 7 shows DH and En results of numerical experiments for the cases $Q_t = -1$ (Fig. 7a), $Q_t = -2$ (Fig. 7b) and $Q_t = -3$ (Fig. 7c). As mentioned above, the TS, when interacting with the DW, breaks up into $Q_t$ pairs of LP moving along the DW plane. At the second stage of the experiments, similarly to the previous cases, the process of interaction between the TS and the DW was modeled in the RT ($t' \to -t$). Thus, models were obtained that describe the formation of TS of the form (5) in the DW (7) plane with their subsequent emission by DW (Fig. 8).

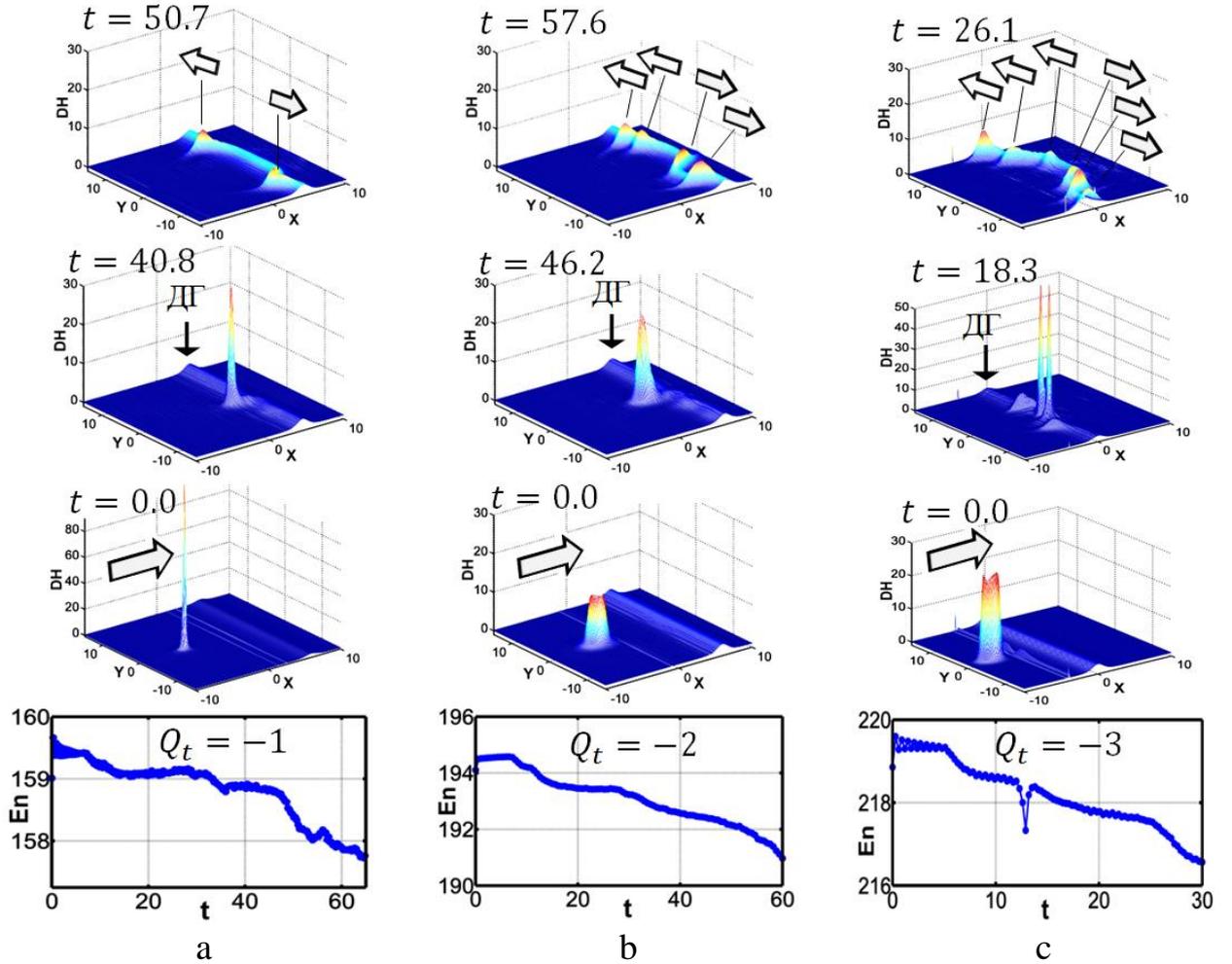

**Figure 7.** The energy density of the DH and En of the processes of collision and annihilation of TS (5) and the DW (7) in ordinary time – t at: $Q_t = -1$ (a); $Q_t = -2$ (b); $Q_t = -3$ (c).

On Fig. 8a shows the evolution of the system (DH and En) consisting of two LP located in the DW plane at a finite distance from each other. This system at $t' = 0$ is equivalent to the state of the model shown in Fig. 6a at $t = 50.7$. As a result of a series of numerical experiments, a model was obtained that describes the processes of formation of a topological perturbation in the DW (7) plane when two localized perturbations are combined. Analysis of the isospin structure of the generated topological perturbation shows its complete identity with the isospin structure of the topological vortex (5) at $Q_t = -1$. In this case, the formed topological vortex has sufficient energy to separate from the DW (see Fig. 8a, at $t' \to 60$).

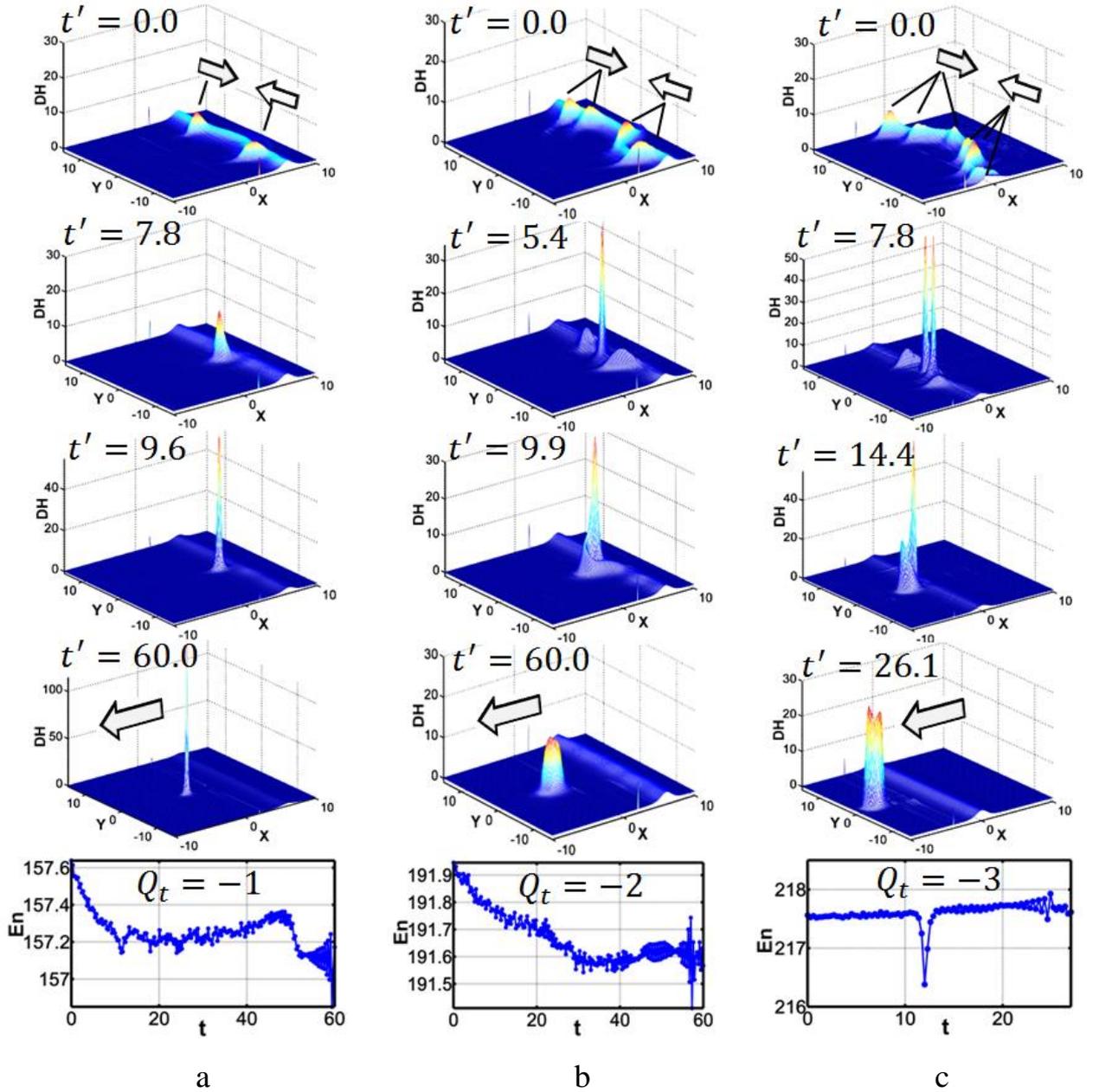

**Figure 8.** The energy density of the DH and En of the processes of collision and annihilation of TS (5) and the DW (7) reversed time – ($t' \to -t$) at: $Q_t = -1$ (a); $Q_t = -2$ (b); $Q_t = -3$ (c).

On Fig. 8b and Fig. 8c shows similar experiments on the formation of topological vortices (5) on the DW plane (7) in the cases $Q_t = -2$ and $Q_t = -3$. As can be seen from these illustrations, in both cases, the manifestation of the T-invariance property of the system of interacting topological fields (5) and (7) is observed with a sufficiently high accuracy. The energy loss for systems obtained by time reversal is negligible: $\text{En}_{loss}(Q_{t'}) \to 0$.

## 5. Results and Discussion

The studied topological solitons are quasiparticles and represent spatial distributions of spins (magnetization directions), which cannot be transferred by continuous transformation into a homogeneous (vacuum) state corresponding to the ground state of the field. It should also be noted that the models obtained are not exact analogues of the results of practical experiments, since the objects of the studies presented are two-dimensional. Accordingly, our results may differ to some extent from the results of practical experiments [18].

Note also that in experiments on the interaction of the TS in ordinary time ($t$) there is a release of excess energy in the form of linear radiation waves (until the formation of an exact solution), which are absorbed by special boundary conditions at the edges $L(x,y)$. Further, when modeling this process in the RT ($t' \to -t$), the evolution of the numerical system consisting of perturbation waves as close as possible to their analytical form takes place. Thus, as shown in the previous part, the T-symmetry property of the models under study is observed with a fairly high accuracy: $\text{En}_{loss} \to 0$.

## 6. Conclusions

Thus, the properties of the T-invariance of a supersymmetric O(3) NSM are considered by the example of the processes of interaction, decay, and annihilation of its topological soliton solutions. The study was carried out by numerical simulation methods based on standard difference equations and a specially developed algorithm using the properties of a stereographic projection. For evolutionary models of head-on collisions and stage-by-stage annihilation of topological vortices, the time reversal operation ($t' \to -t$) is applied. Models are obtained that describe the process of combining radiation waves and the formation of the initial state of interacting topological vortices. The T-invariance property of the studied (2+1)-dimensional O(3) NSM is confirmed and its accuracy in describing the nonlinear dynamics of localized topological perturbations is shown. A package of computer programs has been developed that makes it possible to study the evolution of interacting localized solutions of field theory models in reversed time.

The obtained results show that the developed research method allows one to experimentally restore the initial state of the field, single and interacting localized solutions of nonlinear media controlled by NSM.

**Funding**

The work was carried out as part of the implementation of the research plan of the Department of Nanomaterials and Nanotechnologies of the S.U.Umarov Physical–Technical Institute of the National Academy of Sciences of Tajikistan, grant No. 0119TJ00994.

**References**

[1]. Marion J.B. Physics and the Physical Universe. NY: John Wiley & Sons, Inc., 1971. 712 p.
[2]. Polyakov A.M. Interaction of Goldstone particles in two dimensions. Applications to ferromagnets and massive Yang-Mills fields. Phys. Lett., 1975. V. 59B. № 1. P. 79–81.
[3]. Vainshtein A.I., Zakharov V.I., Novikov V.A., Shifman M.A. Two-dimensional sigma models. Models of nonperturbative effects of quantum chromodynamics // FECHAYA, 1986. V. 17. No. 3. P. 472–545.
[4]. Derode A., Roux Ph., Fink M. Robust Acoustic Time Reversal with High-Order Multiple Scattering // Physical Review Letters, 1995. V. 75. № 23. P. 4206–4210. doi: 10.1103/PhysRevLett.75.4206

[5]. Chabchoub A., Fink M. Time-Reversal Generation of Rogue Waves // Physical Review Letters, 2014. V. 112. P. 124101-1–124101-5. doi: 10.1103/PhysRevLett.112.124101

[6]. Ogawa K., Tamate Sh., Nakanishi T., Kobayashi H., Kitano M. Classical realization of dispersion cancellation by time-reversal method // Physical Review A, 2015. V. 91. P. 013846-1–013846-6. doi: 10.1103/PhysRevA.91.013846

[7]. Melkaer J.M., Wu Z., Braun G.M. Time-reversal-invariant topological superfluids in Bose-Fermi mixtures // Physical Review A, 2017. V. 96. P. 033605-1–033605-6. doi: 10.1103/PhysRevA.96.033605

[8]. Sounas D.L., Alu A. Time-Reversal Symmetry Bounds on the Electromagnetic Response of Asymmetric Structures // Physical Review Letters, 2017. V. 118. P. 154302-1–154302-6. doi: 10.1103/PhysRevLett.118.154302

[9]. Wang Ch., Levin M. Anomaly Indicators for Time-Reversal Symmetric Topological Orders // Physical Review Letters, 2017. V. 119. P. 136801-1–136801-5. doi: 10.1103/PhysRevLett.119.136801

[10]. Tachikawa Y., Yonekura K. Derivation of the Time-Reversal Anomaly for (2 + 1)-Dimensional Topological Phases // Physical Review Letters, 2017. V. 119. P. 111603-1–111603-5. doi: 10.1103/PhysRevLett.119.111603

[11]. Ningyuan J., Schine N., Georgakopoulos A., Ryou A., Sommer A., Simon J. Photons and polaritons in a broken-time-reversal nonplanar resonator // Physical Review A, 2018. V. 97. P. 013802-1–013802-10. doi: 10.1103/PhysRevA.97.013802

[12]. Kozyryev I., Hutzler N.R. Precision Measurement of Time-Reversal Symmetry Violation with Laser-Cooled Polyatomic Molecules // Physical Review Letters, 2017. V. 119. P. 133002-1–133002-6. doi: 10.1103/PhysRevLett.119.133002

[13]. Leutenegger T., Dual J. Detection of defects in cylindrical structures using a time reverse method and a finite-difference approach // Ultrasonics, 2002. V. 40. P. 721–725. doi: 10.1016/S0041-624X(02)00200-7

[14]. Huang R. Quantum Field Theory. From Operators to Path Integrals. New York: John Wiley & Sons, Inc., 1998. 426 p.

[15]. Peskin M.E., Schroeder D.V. An Introduction to Quantum Field Theory. San Francisco: Addison-Wesley Publ., 1995. 842 p.

[16]. Kudryavtsev A., Piette B., Zakrzewski W. Skyrmions and domain walls in (2+1) dimensions // Nonlinearity, 1998. V. 11. P. 783–795.

[17]. Muminov Kh.Kh. Multidimensional dynamic topological solitons in a nonlinear anisotropic sigma model // DAN RT. 2002. V. 45. No. 10. S. 28–36.

[18]. Muminov Kh.Kh., Shokirov F.Sh. Mathematical modeling of nonlinear dynamic systems of quantum field theory. Novosibirsk: Publishing House of SO RAN, 2017. 375 p.

[19]. Muminov Kh.Kh., Shokirov F.Sh. Interaction and decay of two-dimensional topological O(3) solitons of a vector nonlinear sigma model // DAN RT. 2011. V. 54. No. 2. S. 110–114.

[20]. Belavin A.A., Polyakov A.M. Metastable states of a two-dimensional isotropic ferromagnet, JETP Letters, 1975, vol. 22, no. 10, pp. 503–506.

[21]. Kudryashov N.A. Methods of nonlinear mathematical physics. M.: MEPhI, 2008. 352 p.

[22]. Kovalev S.A. Vortex structure of magnetic solitons // Physics of low temperatures. 2017. V. 43. No. 2. S. 334–346.

[23]. Muminov Kh.Kh., Shokirov F.Sh. Isospin dynamics of topological vortices // DAN RT. 2016. V. 59. No. 7–8. pp. 320–326.
[24]. Muminov Kh.Kh., Shokirov F.Sh. Dynamics of interaction of topological vortices with a domain wall in a (2+1)-dimensional nonlinear sigma model // DAN RT. 2015. V. 58. No. 4. S. 302–308.
[25]. Samarsky A.A. Theory of difference schemes. M.: Nauka, 1989. 616 p.
[26]. Muminov Kh.Kh., Shokirov F.Sh. Dynamics of interactions of two-dimensional topological solitons in O(3) of a nonlinear vector sigma model // DAN RT, 2010, v.53, no. 9, pp. 679 – 684.